\documentclass[aps,pra,reprint,showpacs,amsmath,amssymb,amsfonts]{revtex4-1}
\pdfoutput=1
\usepackage{blindtext}
\usepackage{graphicx}
\usepackage{fullpage}
\usepackage{dcolumn}
\usepackage{hyperref}
\usepackage{tabularx}

\begin{document}
\title{Characterization of Electron Pair Velocity in
YBa$_{2}$Cu$_{3}$O$_{7-\textit{$\delta $}}$ Thin
Films}
\author{Ronald Gamble, Jr}
\email{rsgamble@aggies.ncat.edu}
\author{K.M.Flurchick}
\affiliation{Department of Computational Sciences \& Engineering}
\author{Abebe Kebede}
\affiliation{Department of Physics\\North Carolina A \& T State
University\\Greensboro, NC, 27401}

\begin{abstract}
The superconducting phase transition in YBa$_{2}$Cu$_{3}$O$_{7-\textit{$\delta
$}}$(YBCO) thin film samples doped with non-superconducting nanodot impurities
of CeO$_{2}$ are the focus of recent high-temperature superconductor studies.
Non-superconducting holes of the superconducting lattice induce a bound-state
of circulating paired electrons. This creates a magnetic flux vortex state.
Examining the flow of free-electrons shows that these quantized magnetic flux vortices arrange themselves in a self-assembled lattice. 
The nanodots serve to present structural properties to constrict the "creep" of these flux vorticies under a field response in the form of 
a pinning-force enhancing the critical current density after phase transition.
In this work, a model for characterizing the superconducting phase by the work
done on electron pairs and chemical potential, following the well-known theories
of Superconductivity (Bardeen-Cooper-Scheifer \& Ginzburg-Landau), is
formulated and tested.
A solution to the expression for the magnetic flux, zero net force and pair velocity will generate 
a setting for the optimal deposition parameters of number density, growth geometry and mass density of these nanodot structures.
\end{abstract}

\maketitle
\section{Introduction}

High-Temperature Superconductivity (HTS) is the coherent ordering of macroscopic
quantum states for valence band electrons. HTS occurs at temperatures higher
than that of liquid Nitrogen ($~$77K). These coherent or paired valence band
electrons, referred to as \textit{Cooper Pairs}\cite{bcs} account for the
persistent electrical currents within the (a,b)-surfaces of the lattice structure for the conducting material. Modifications
with non-superconducting properties can serve to restrict the motion of magnetic
flux vorticies via a pinning-force as well as enhance the critical current
density.
High-Temperature Superconductivity is a
very promising field of study given it{'}s marriage between macroscopic quantum phenomenology and emerging
technologies on the mesoscale and nanoscale. A number of
complications arise experimentally; including the cost of creating each sample
superconductor (e.g. laser ablation, doping agents, vacuum environments, etc.)
and the current limitations (i.e.
physical properties measurements, SQUID, and again laser ablation, etc.). These
complications generate difficulties when attempting to characterize the
superconducting compound. Multiple techniques are employed to overcome some of
these difficulties, using a variety of sample growth methods (i.e. single $\&$
multi-layer growth modes), introducing structual impurities into the
superconductor that strengthens the overall electrodynamics of the system and
other experimental techniques to help characterize the sample.\\

The work described here extends the understanding of the characterization of
superconducting compounds in terms of the lattice modifications, nanodot
impurities, and applying the modifactions as a basis for further characterizing
the superconducting sample in terms of the interaction between the nanodot and
the superconducting electrons. Using the description \cite{bcs} and the
derived thermodynamic formulation of the superconducting system by Ginzburg and
Landau \cite{Ginz}, a new theoretical approximation of the electron pair
velocity is presented. A formulation of the new model for the system is given by
a variation in the electron pair velocity from a ficticious force generated by
the presence of a nanodot. The model is tested using the results from
(T.Haywood et al.)\cite{Haywood}.
A comparison is made between experimental and theoretical velocity calculations
using growth geometry and total chemical potential. This referenced work
contained a very good basis in regards to having two different \(\text{CeO}_2\)
deposition methods, substrate modifications and multilayer growth, while using
the stable Volmer-Weber growth mode introducing 3-dimensional surface
modifactions.

 This model gives insight into how the current density for a doped
 high-temperature superconductor will be modified and tuned based on the dynamics and density of the nanodots themselves.
Electron pair velocities can be calculated using the current density,
collective charge of the superconducting pair and the number density of the
superfluid from the referenced work above. \cite{Haywood,beasley,kondo}.

\section{Magnetic Flux and Critical Current Density Distributions}

It is known that magnetic flux through a ring of supercurrent will become
quantized thus, creating a magnetic flux vortex from the torsion effects of the
supercurrents\cite{abri}. The normal zones of Cerium Oxide
\(\left(\text{CeO}_2\right.\)) deposited onto the thin film samples through
laser ablation serve as field penetration sites permittes magnetic flux
lines to pass through the sample, in a {``}swiss cheese{''}-like structure. The
creation of these magnetic flux vortices introduces a vortex state in the
sample, existing between the lower and upper critical field limits where
\(H_{\text{lower}}<H_{\text{vortex}}<H_{\text{upper}}\).
The Vortex state of YBCO follows the Abrikosov Vortex lattice theory for the
anisotropic surfaces of type-II superconductors \cite{abri}.
Focusing on the magnetic flux penetrating the sample we look into how this
combination of flux and lattice hole effects the flow of paired electrons,
inertially. A solution to the expression for the magnetic flux, zero net force
and pair velocity will generate a setting for the optimal deposition parameters
of number density, growth geometry and mass density of these nanodot structures.
From the dimensional analysis describing magnetic flux, one can derive a
relationship between  work and current density. The standard unit of measure for
magnetic flux is normally a Weber (Wb) or a Tesla square meter \(\left(T\cdot
m^2\right)\). These units can be simplified into fundamental terms with respect
to the M$\cdot $K$\cdot $S system of measure \(\text{Wb} = T\cdot
m^2=\text{Kg}\cdot m^2/{s^2\cdot A}\). Now that the magnetic flux is recast in
to standard units of length, mass and time, an expression describing the same
physical action will be constructed that corresponds to the magnetic flux units
of measure. Including the magnetic field effects with the current density, the
velocity of each pair is now expressed in terms of the magnetic vector potential (\pmb{\textit{A}}) and the quantum mechanical
representation of the potential energy of the state \((\hbar
 \nabla \phi )\)\cite{bcs}:
\begin{equation}
 \pmb{V}_*=\frac{1}{m_*}\left(\hbar \nabla \phi
 -\frac{e_*}{c}\pmb{A}\right)\label{vel1}
\end{equation}

With the inclusion of the electron pair mass and quantization via
($\hbar\nabla\phi$) in equation (\ref{vel1}) for the canonical velocity of the
electron pair suggests that the quantum mechanical operations for this coherent state of
electrons is of a macroscopic nature, corresponding to an inertial
response with respect to the mass term.

The expectation values, probability amplitude, and average densities are
associated to observable values and not probabilistic in nature. This
macroscopic quantum mechanical expression for the canonical velocity of paired
electrons gives rise to the \textit{inertial} dynamics of the pair themselves.
This states that the critical current density of the pairs, and fundamentally
the pair velocity, is reactive to some external inertial force acting on the
center of mass of the pair. The critical current density is stated as
\cite{bcs}:
\begin{eqnarray}
 \pmb{J}_* &=& \frac{e_*|\psi |^2}{m_*} \left(\hbar \nabla \phi
 -\frac{e_*}{c}\pmb{A}\right)  \\
 &=&\frac{e_*n_*}{m_*}\left(\hbar \nabla \phi
 -\frac{e_*}{c}\pmb{A}\right)\label{curr1} 
 \end{eqnarray}

with \(\left(e_*\right)\) the charge, \(\left(m_*\right.\)) the mass, and
\(\left|\psi |^2\equiv n_*\right.\) is the number density of the electron pairs;
and (c) is the speed of light. A modification to this distribution of the
supercurrent density due to the presence of nanodots suggests that the nanodots,
normal zones of the lattice. Introducing a pseudo-potnetial well that the paired
electrons fall into will thus change the supercurrent density\cite{kondo}. From
the analysis of magnetic flux the {``}electronic hole{''} that is made from the presence of nanodots serves as an enclosed area that can be
determined. The superconducting electrons can circumvent the enclosed surface,
strengthening the supercurrent density.
\begin{equation}
\pmb{V}_*=\frac{\pmb{J}_*}{n_*e_*}\label{vel2}
\end{equation}
enclosing a single fluxon\cite{tozan}.\\

\section{Chemical Potential of the Normal-Superconducting State Interaction}

To modify the description of the thermodynamic dependence of the critical
current density and the magnetic flux threading the superconductor in the
presence of nanodots, a reformulation of the fundamental free-energy expression
is needed\cite{Ginz,beasley}:
\begin{eqnarray}
 dF=-SdT - \underset{k}\sum  \pmb{Q}_k{}^{(r)}
 d\pmb{\text{r}}_k{}^{(r)}+\underset{j}\sum\mu _j\pmb{d}N_j\label{gl1}
\end{eqnarray}
The free-energy expression from the Ginzburg-Landau theory\cite{Ginz,beasley}
says:
\begin{equation}
 F_{G-L}(r) =F_{\text{grad}}(r)
 +F_{\text{Lattice}}(r)+U_{\text{magnetic}}(r)\label{gl2}
\end{equation}
Here the standard entropy (S) and temperature (T) terms are held constant for
this thermodynamic state. With a force and coordinate in thermodynamic state (r),
\(\sum _k\pmb{Q}_k{}^{(r)}\pmb{\text{dr}}_k{}^{(r)}\) is the amount of work from
the nanodot interacting with the system of electron pairs and
$\sum_j\mu _j\pmb{d}N_j$ is the chemical potential with respect to the
number density (N). Utilizing the work-done on a system of particles combined
with the quantized magnetic flux quasiparticles called Fluxon\cite{abri}, one can formulate
a description of the free-energy interaction of these Fluxon with the
supercurrent density surrounding them in terms of the chemical potential and
nanodot number density. For the system acted upon by an interacting potential,
the free-energy is:
\begin{eqnarray}
dF=dU-SdT+\sum _j\mu _jdN_j\label{gl3}
\end{eqnarray}
where,
\begin{eqnarray}
 dU=TdS-\pmb{r}_kd\pmb{\text{$\chi$}}_k+dF_{G-L}\label{gl4}
\end{eqnarray}

With respect to the thermodynamics of the superconducting sample the chemical
potential of all interacting particles and quasiparticles must be included. The
number density of interacting particles and each of their chemical, or
electro-chemical, potentials can alter the dynamics of the thermodynamic system.
The energy of the paired
electrons is just simply their electro-chemical potential in this quantum limit.
Considering interactions that occur the total chemical
potential is:
\begin{eqnarray}\nonumber
 \mu _{\text{tot}}&=&\underset{i}\sum\frac{\partial}{\partial
 N_i}\left[\frac{1}{\beta }\left(\log (z)+\text{$\beta $U}_i\right)-
 \underset{k}\sum \pmb{Q}_k{}^{(r)} d\pmb{\text{$\chi $}}_k{}^{(r)}\right]\\
 &+&\underset{j}\sum\frac{\partial }{\partial N_j}\left[\phi
 _*\right]+\underset{i}\sum\frac{\partial }{\partial N_i}(m\cdot B)\label{mu1}
\end{eqnarray}

Where $\beta, z, U_i$ are the lattice parameters, with  the
magnetic influence ($m\cdot B$), paired electron potential ($\phi_*$), and
fluxon/nanodot extent ($\pmb{\chi}_k$). The total chemical potential of the entire system suggests
that there are other quasiparticles at play interacting with the paired
electrons comprising up the supercurrent. Simplifying this total chemical
potential in equation (\ref{mu1}) we have \(\mu _{\text{tot}}=\mu _k+\mu _*\).
Where \(\mu _k\) is the chemical potential of the nanodot and \(\mu _*=\mu _e+e_*\phi
_*\) is the electro-chemical potential for the electron pair in terms of the
thermodynamic chemical potential of the pair and the electrostatic
potential for charged particles. 

 Considering the dimensions of the
nanodots as \(\left(\pmb{\chi} _{\beta }, \pmb{\chi} _{\gamma }\right)\), where
these are the respective diameter (with plane-polar symmetry) and height of the
nanodots, we can assume that the geometry of the nanodots follow that of a
spheroid.

\begin{table}[h]
\caption{\label{tab:table2}
Substrate Modification: THA (10 pulses), THB (30 pulses) Multilayer: THA1 (10
pulses), THB1 (30 pulses)\cite{Haywood}}
\begin{ruledtabular}
\begin{tabular}{c c c}
Samples & Approx. Diameter & Approx. Height\\
 THA \& THA1 & 4.0-6.0\text{nm} & 1.77\text{nm} \\
 THB \& THB1 & 4.0-6.0\text{nm} & 4.0\text{nm} \\
\end{tabular}
\end{ruledtabular}
\end{table}

The average volume of each Cerium Oxide nanodots can be calculated using the
following equation for a spheroid with plane-polar symmetry,
\begin{eqnarray}\nonumber
 V_{\text{\textit{spheriod}}}&=&\frac{4}{3}\pi (a)^2c\\
 &=&\frac{4}{3}\pi \left(\frac{\tilde{\chi _{\beta
 }}}{2}\right){}^2\tilde{\chi }_{\gamma }\label{vol1}
\end{eqnarray}

\begin{table}[h]
\caption{\label{tab:table2}
Substrate Modification: THA (10 pulses), THB (30 pulses) Multilayer: THA1 (10
pulses), THB1 (30 pulses)\cite{Haywood}}
\begin{ruledtabular}
\begin{tabular}{ccc}
Samples & Approx. Radius & Approx. Volume\\
 THA \& THA1 & 2.0\text{nm} & 28.48377$\text{nm}^3$ \\
 & 2.5\text{nm} & 44.50589$\text{nm}^3$ \\
 & 3.0\text{nm} & 64.08849$\text{nm}^3$ \\
 THB \& THB1 & 2.0\text{nm} & 67.02064$\text{nm}^3$ \\
 & 2.5\text{nm} & 107.71975$\text{nm}^3$ \\
 & 3.0\text{nm} & 150.79644$\text{nm}^3$ \\
\end{tabular}
\end{ruledtabular}
\end{table}

Cerium Oxide with a mass density of $\sim$
7.2148$\times10^{-22} \text{g/nm}^3$ gives an average mass of the nanodots based
on the density of Cerium Oxide and the average volume of the nanodots. With
this property we can calculate approximate masses for the 10 pulse and 30
pulse Cerium Oxide volumes, respectively.\\

\section{Inertial Response of the Electron Pair}

\begin{table}[h]
\caption{\label{tab:table2}%
Approximate Nanodot Mass}
\begin{ruledtabular}
\begin{tabular}{cc}
Mass ($10^{-20}$g) at 10 pulses & Mass($10^{-20}$g) at 30 pulses\\
 2.055 & 4.83542 \\
 3.211 & 7.77179 \\
 4.624 & 10.87971 \\
 (Average Mass) 3.29667 & (\text{Average} \text{Mass})7.82897 \\
\end{tabular}
\end{ruledtabular}
\end{table}

\begin{figure}
\includegraphics[height=5cm,width=\linewidth]{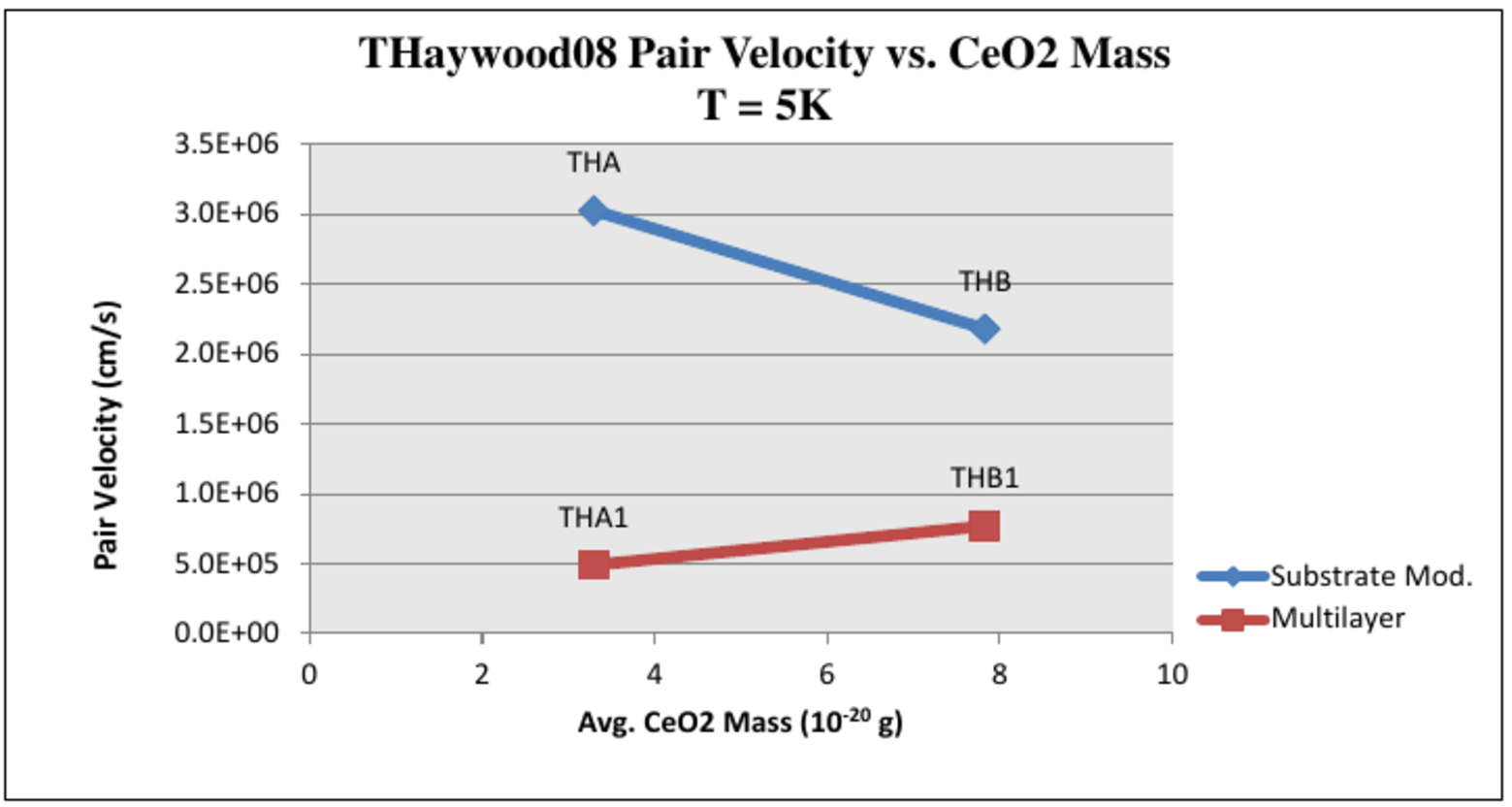}
\caption{Averaged electron pair velocity correlation with average
$\text{CeO}_2$ mass at 5K\cite{Haywood}}

\includegraphics[height=7cm,width=\linewidth]{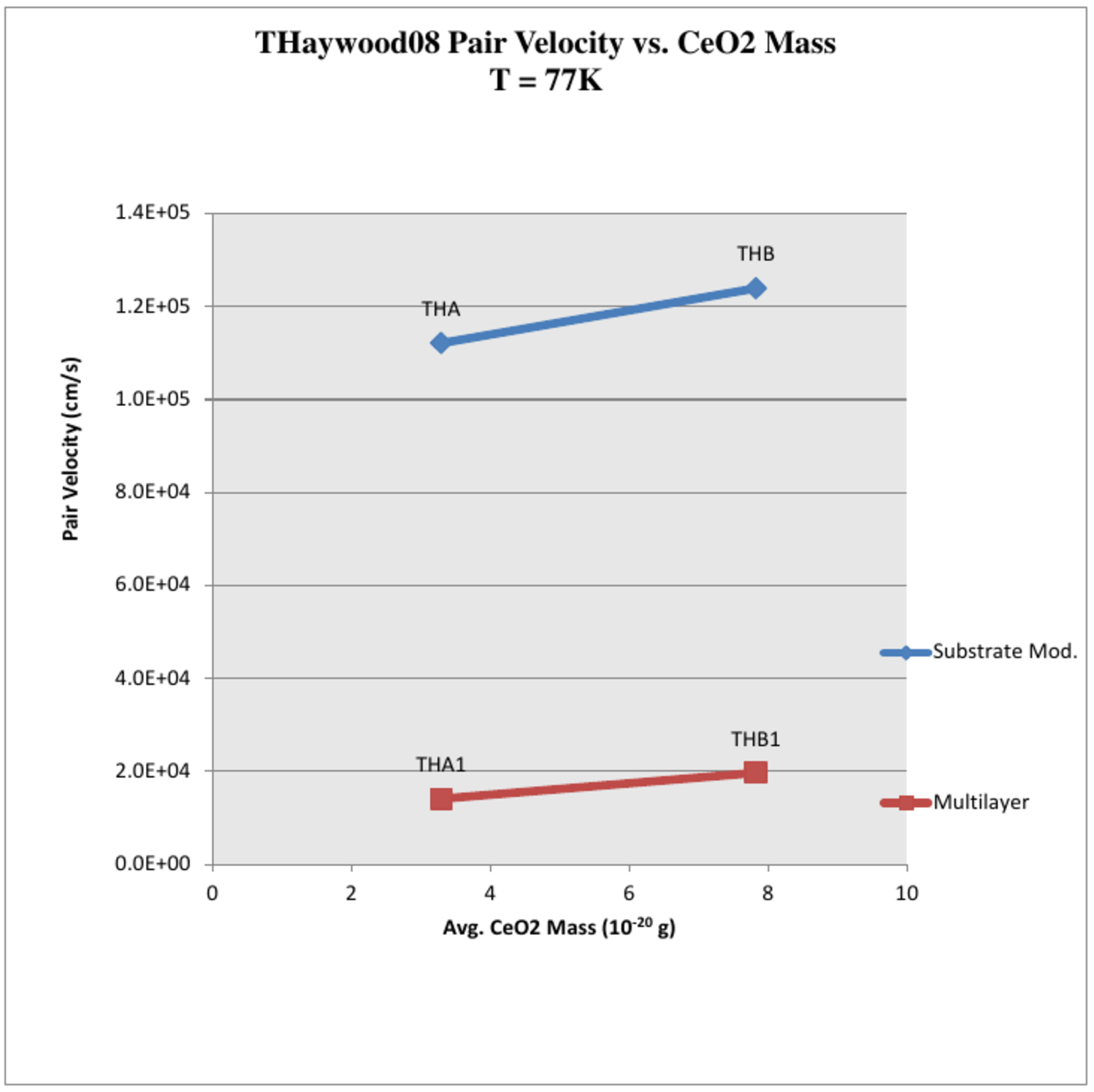}
\caption{Averaged electron pair velocity correlation with average $\text{CeO}_2$
mass at 77K\cite{Haywood}}
\end{figure}

An expression describing the magnetic flux through the superconducting-normal
lattice zones can be given as, using the equation above:
\begin{eqnarray}\nonumber
 \mu _{\text{tot}}&=&\underset{i}\sum\frac{\partial}{\partial
 N_i}\left[\frac{1}{\beta }\left(\log (z)+\text{$\beta $U}_i\right)-
 \underset{k}\sum \pmb{Q}_k{}^{(r)} d\pmb{\text{$\chi $}}_k{}^{(r)}\right]\\
 &+&\underset{j}\sum\frac{\partial }{\partial N_j}\left[\phi
 _*\right]+\underset{i}\sum\frac{\partial }{\partial N_i}(m\cdot \pmb{B})
\end{eqnarray}
\\
This total chemical potential simplifies to \(\mu _{\text{tot}}=\mu _k+\mu _*\). 
\begin{equation}
 \mu _*=\mu _e+e_*\phi _*=-E \\
\end{equation}
and
\begin{equation}
\mu _{\text{k}}=\underset{i}\sum\frac{\partial}{\partial
 N_i}\left[\frac{1}{\beta }\left(\log (z)+\text{$\beta $U}_i\right)-
 \underset{k}\sum \pmb{Q}_k{}^{(r)} d\pmb{\text{$\chi $}}_k{}^{(r)}\right]
\end{equation}

Since the lattice structure of YBCO is periodic with respect to the electron pairs with temperature equal to zero, an approximation for the chemical
potential governing the nanodots can be made in the form of the  work:
\begin{equation}
 \mu _k= -\frac{\partial }{\partial N_i}
 \left[\underset{k}\sum \pmb{Q}_k{}^{(r)} \pmb{\text{d$\chi
 $}}_k{}^{(r)}\right]\label{mu2}
\end{equation}

Equation (\ref{mu2}) neglects the magnetic dipole moment and field because of
the hole like behavior of the nanodots. For the nanodots acting as electron
holes one can approximate these as a neutral mass. Utilizing the work-done from
the perspective of the nanodots is an unconventional choice. From the BCS theory
the paired electrons have a velocity, refering to eqaution (\ref{vel1}) above.
With the inclusion of the mass, the paired electrons respond to a force giving
an acceleration, \(\frac{\pmb{\nabla} \mu
_*}{m_*}=\frac{d}{dt}\left[\pmb{v}_*\right]\).
The electron pairs respond to an inertial force. We see that it is obvious
in these units that magnetic flux is merely the amount of work per current. A
net force can be expressed from the interaction of the electron pair and the
nanodots:
\begin{subequations}
\begin{equation}
 \pmb{F}_{*, \text{net}}=-\pmb{\nabla }\left(\mu _k+\mu _*\right)
 \end{equation}
where, $+\pmb{\nabla }\mu _k=-(\pmb{\nabla }\mu _*)$ and the net force can be
rexpressed as:
 \begin{eqnarray} 
 \pmb{F}_{*, \text{net}}&=&-\pmb{\nabla }(\pmb{Q}_k{}^{(r)}d\pmb{\chi
 }_k{}^{(r)})+\left(-\pmb{\nabla }\left(\mu _e+e_*\phi \right)\right)
 \\
 &=&-\pmb{\nabla }(\pmb{Q}_k{}^{(r)}\cdot \pmb{\xi
 _*})+\left(-\pmb{\nabla }\left(\mu _e+e_*\phi \right)\right)
 \end{eqnarray}
 \end{subequations}

Here the work-done is in a thermodynamic energy state (\textbf{\textit{r}})
operating within the canonical momentum space of the system, and as usual the
electrochemical potential arises for the electron pairs\cite{groff,henley}. Like
all systems in equilibrium, this net force must equal to zero satisfying the conservation of
energy and momentum of the interaction per the 2nd Law of Thermodynamics. Using
this, the magnetic flux can be expressed as
\begin{eqnarray}
 \Phi _k=\frac{\left(\pmb{Q}_k\cdot \pmb{\xi }_*\right)}{\pmb{J}_*\cdot\left(\pi \left(\frac{\pmb{\chi }_{\beta }}{2}\right){}^2\right)}+\Phi_0\label{flux1}
\end{eqnarray}

Magnetic flux is in terms of the current density, an equivalent inertial force
and the coherence length describing the size of the electron pairs related to
the displacement the pairs should experience from the work. As can be seen,
\(\left(\pmb{J}_*\right)\) is the respective supercurrent density of the sample
at a specific temperature, (\(\pi \left(\frac{\pmb{\chi }_{\beta
}}{2}\right){}^2\)) is the cross-sectional area of the nanodots keeping the
radial symmetry of the geometry. While \(\left(\pmb{\xi }_*\right.\)) is the
characteristic superconducting coherence length, \(\left(\pmb{Q}_k\right)\) the
force induced by the magnetic flux on a charged particle, and \(\left(\pmb{\Phi
}_0\right)\) the quantum of magnetic flux (Fluxon, 2.0678
\textit{x}\(10^{-15}\text{Wb}\)). This force arises from the potential energy
that the nanodot creates on the surface of the superconducting state in momentum
space. Without exploring the entire effective field theory for superconductivity
only an approximation of the characterized average velocity of paired electrons
can be made. Using the fundemental laws that governs this electromagnetic
interaction we can approximate or generalize the expression for current density
in equation (\ref{curr1}) to be \(\pmb{J_*}=n_*e_*\pmb{v}_*\simeq
\frac{\pmb{I}}{\pi \pmb{s}^2}\).
For an approximation of a simple, homogeneous applied magnetic field \(B_a\)
(assuming no applied magnetic field excitaions) with magnetic flux ($\Phi $)
through an enclosed current carrying loop of radius S, we can use the solution
of
\begin{eqnarray}
 \Phi &\approx&\oint\left(B_a\cdot\hat{n}\right)ds\\
 &\cong& B_a\text{$\pi $s}^2 \nonumber\\
 &\approx& B_a\pi \left(\frac{\chi _{\beta }}{2}\right){}^2 \nonumber
\end{eqnarray}
From here we can solve for the current density and then the velocity of the
electron pairs from equation (\ref{flux1}).
\begin{eqnarray}
 J_*&\simeq& \frac{\chi_{\beta}\left(-\nabla\left(Q_k\cdot\xi
 _*\right)\right) +\Phi_0}{\Phi_k\sigma}\label{curr2}
 \\
 &\simeq& \frac{\left(Q_k\cdot \xi _*\right)}{B_a\pi \left(\frac{\chi _{\beta
 }}{2}\right){}^2\left(\pi \left(\frac{\chi _{\beta
 }}{2}\right){}^2\right)}\label{curr3}
\end{eqnarray}
Equations (\ref{curr2}) and (\ref{curr3}) makes this approximation in terms of
the induced force portrayed by the Lorentz force. The electric field contribution is negligible due to the macroscopic electrodynamics explained through the
London theory.
\newpage

\begin{eqnarray}
 \omega _*\approx\frac{\left(-\nabla \left(e_*v_*B_a\text{sin$\gamma $}\right)\cdot \xi _*\right)}{\pi ^2\left(\frac{\chi _{\beta }}{2}\right){}^4\cdot
B_a}\frac{1}{n_*e_*}\label{vel3}
\end{eqnarray}
This approximation in equation (\ref{vel3}) unfortunately gives a fairly wide
range of percent error, percent error $\approx (0.253 - 7.895) \times 100\%$, due to the lack of a
temperature dependance on the supercurrent density approximation from the
quantum mechanical field theory. Figures (3) and (4) below show that these
theoretical values closely equal in orders of magnitude to the experimental values with some
percent error in calculation.

\begin{figure}[h]
\includegraphics[height=6.7cm,width=\linewidth]{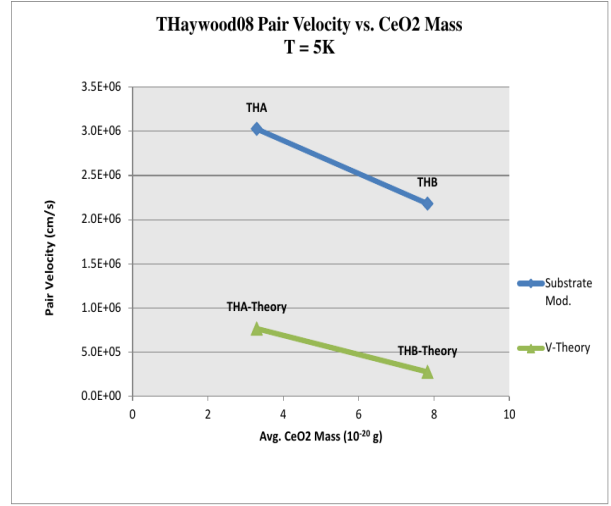}
\caption{Averaged theoretical Electron pair velocity correlation with average
$\text{CeO}_2$ mass at 5K [Substrate Modification]}
\end{figure}
\begin{figure}[h]
\includegraphics[height=6.7cm,width=\linewidth]{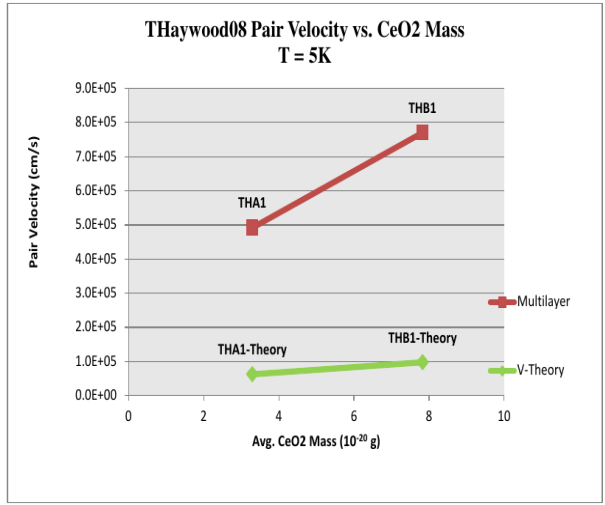}
\caption{Averaged theoretical Electron pair velocity correlation with average
$\text{CeO}_2$ mass at 5K [Multilayer Growth Modification]}
\end{figure}

The error in the multilayer growth method is higher due to the approximation
methods used for the \(\text{CeO}_2\) nanodots.
We can see that there is a stronger relationship between the nanodots and the
paired electrons in terms of their velocity. This suggests that theses magnetic
flux vortices penetrating at the normal zones of the sample offer more than
expected of them. These normal zones may offer an optimization to the
superconducting sample instead of a defect in structure.
\newpage

\section{Conclusion}

This description of the variation of the superconducting electron pair velocity
is incomplete, however it demonstrates a mechanism in terms of further
characterizing high-temperature superconductors. Characterization in terms of
the respective nanodot densities and geometries deem critical to the enhancement
of supercurrent density.
The chemical potential and work-done from a constant thermodynamic energy state
offer a method of describing an induced force that arises from lattice
modifications via single and multi-layer Volmer-Weber growth modes. Equation
(\ref{vel3}) provides an expression of the modified average electron pair
velocity which can be viewed as a \textit{predicted} quantity. The term
($V_*$) in the expression serves as a the velocity of a control sample
with unmodified lattice structure (absence of nanodots). While ($\omega_*$) is
the predicted modified velocity of the superconducting electrons under an
applied magnetic field ($B_a$) with nanodot diameter ($\chi_\beta$). Using this
expression one can calculate a predicted average velocity and thus
supercurrent density at (T = 5K) before deposition of any lattice
modifications (within a 25 percent error).

Next steps will inlcude a richer description of the temperature dependence of
the superconducting state to allow for a scalable description of the velocity
with respect to the state's effective temperature. Further correlating this
description with the Abrikosov Vortex lattice theory [abrikosov, 1957] and
continued research on the subject matter will generate interesting results to
the study of theses magnetic singularities (Fluxon) in high-temperature
superconductors. The overall effective field theory governing
this interaction is to be explored in greater detail.

\bibliographystyle{aipauth4-1} % Tell bibtex which bibliography style
% to use
\bibliography{VORTEX_master.bib} % Tell bibtex which .bib file to use (this one
% is some example file in TexLive's file tree)

\end{document}